\documentclass[twocolumn,aps,pra,groupedaddress,showpacs]{revtex4}
\usepackage{epsfig,amssymb}
\def\comment#1{}\def\labell#1{\label{#1}}
\begin{document}
%{\scriptsize Eprint: quant-ph/}
%Title of paper
\title{Quantum limits to dynamical evolution}
\author{Vittorio Giovannetti$^1$, Seth Lloyd$^{1,2}$, and Lorenzo
Maccone$^1$}\affiliation{$^1$Massachusetts Institute of Technology --
Research Laboratory of Electronics\\$^2$Massachusetts Institute of
Technology -- Department of Mechanical Engineering\\ 77 Massachusetts
Ave., Cambridge, MA 02139, USA}
%\date{\today}

\begin{abstract}
We establish the minimum time it takes for an initial state of mean
energy $E$ and energy spread $\Delta E$ to move from its initial
configuration by a predetermined amount. Distances in Hilbert space
are estimated by the fidelity between the initial and final state. In
this context, we also show that entanglement is necessary to achieve
the ultimate evolution speed when the energy resources are distributed
among all subsystems.
\end{abstract}
\maketitle

How fast can a quantum system evolve in time, given a certain amount
of energy? If the system is composed of a number of subsystems, is
entanglement a useful resource in speeding up the dynamical evolution? 
To answer the first of these two questions, one typically defines some
characteristic time of the dynamics and studies its connections with
the energy resources of the initial state of the system. Most of the
previous results in this field {\cite{bata,uncer,peres}} trace back to
the time-energy uncertainty relation in the form derived by Mandelstam
and Tamm {\cite{man}}: in this way, the various lifetimes are bounded
by the energy spread $\Delta E$ of the system. More recently, however,
Margolus and Levitin have pointed out that one can relate the
characteristic times of the system also to the average energy $E$ of
the initial state {\cite{margolus}}. In particular, defining the
lifetime of the system as the time it takes for it to evolve to an
orthogonal state, the above results allows one to introduce a {\it
quantum speed limit time} as the minimum possible lifetime for a
system of average energy $E$ and spread $\Delta E$. In {\cite{role}}
we have analyzed such a bound in the case of composite systems
(i.e. systems composed by a collection of subsytems). In this context
we showed that entanglement is a fundamental resource to reach the
quantum speed limit when the energy resources ($E$ or $\Delta E$) are
distributed among the subsystems.

In this paper we extend these results by analyzing what happens when
the quantum speed limit time is generalized redefining it as the
minimum time $t$ it takes for the initial state $\varrho$ to evolve
through a unitary evolution to a state $\varrho(t)$ such that the
fidelity $F(\varrho,\varrho(t))$ of {\cite{fidelity}} is equal to a
given $\epsilon\in[0,1]$. Even though the scenario is more complex
than the case $\epsilon=0$ of {\cite{role}}, also in this case it is
possible to show that entanglement is necessary to achieve speedup of
the dynamics if one wants to share the energy resources among the
subsystems.

In Sect.~\ref{s:riass} we extend the definition of quantum speed limit
time and derive its expression in terms of the energy characteristics
of the initial state, first considering the case of pure states and
then extending the analysis to the more complex case of non-pure
states (Sect.~{\ref{s:qslmix}}). In Sect.~\ref{s:ent} we analyze the
role that entanglement among subsystems plays in achieving the quantum
speed limit. Most of the technical details of the derivations have
been inserted in the appendixes.

% QSL con epsilon
\section{Quantum speed limit}\labell{s:riass}
Since the Hamiltonian $H$ is the generator of the dynamical evolution
and defines the energy of a system, one expects that the energy
characteristics of a state are closely linked to the characteristic
times of its dynamics. In particular, we are interested on how the
mean energy $E$ and the spread $\Delta E$ relate to the maximum
``speed'' the system can sustain in moving away from its initial state
$|\Psi\rangle$.

Take the energy basis expansion of the initial state \begin{eqnarray}
|\Psi\rangle=\sum_nc_n|n\rangle
\;\labell{energ},
\end{eqnarray}
which has average energy $E=\langle\Psi|H|\Psi\rangle$ and spread
$\Delta E=\sqrt{\langle\Psi|(H-E)^2|\Psi\rangle}$. To characterizes
the departure of the system from $|\Psi\rangle$ we can use the
fidelity
\begin{eqnarray} P(t)=\left|\langle\Psi|\Psi(t)\rangle\right|^2=
\left|\sum_n|c_n|^2e^{-iE_nt/\hbar}
\right|^2
\;\labell{pidit},
\end{eqnarray}
where $E_n$ is the energy eigenvalue of the Hamiltonian $H$ relative
to $|n\rangle$. The quantity $P(t)$ is the overlap between the time
evolved state $|\Psi(t)\rangle$ and the initial state
$|\Psi\rangle$. A measure of ``speed of the dynamical evolution'' is
obtained by analyzing how fast $P(t)$ changes in time: e.g. given a
value $\epsilon\in[0,1]$, how long do we have to wait before the state
``rotates'' by an amount $\epsilon$, i.e. before $P(t)=\epsilon$? 
Assuming (without loss of generality) zero ground-state energy
eigenvalue, it is possible to prove that the minimum time at which
this happens is bounded by the quantity\begin{eqnarray} {\cal
T}_\epsilon(E,\Delta
E)\equiv\max\left(\alpha(\epsilon)\frac{\pi\hbar}{2E}\;,\
\beta(\epsilon)\frac{\pi\hbar}{2\Delta E}\right)
\;\labell{newqsl},
\end{eqnarray}
where $\alpha(\epsilon)$ and $\beta(\epsilon)$ are the functions plotted
in Fig.~\ref{f:alpha}.  Of course for $\epsilon=1$ this quantity is
equal to zero: in fact, no time has to pass to obtain $P(t)=1$. On the
other hand, since for $\epsilon=0$ we have
$\alpha(\epsilon)=\beta(\epsilon)=1$, Eq.~(\ref{newqsl}) reduces to the
quantum speed limit time that was defined in {\cite{role,margolus}},
\begin{eqnarray}
{\cal
T}_0(E,\Delta E)\equiv\max\left(\frac{\pi\hbar}{2E}\;,\
\frac{\pi\hbar}{2\Delta E}\right)
\;\labell{qsl},
\end{eqnarray}
which gives the minimum time it takes for a system to evolve to an
orthogonal configuration. In the remainder of the paper, however, with
`quantum speed limit time' we will refer to the generalized version
${\cal T}_\epsilon(E,\Delta E)$ of Eq.~(\ref{newqsl}).

% Da interpola.f
\begin{figure}[hbt]
\begin{center}
\epsfxsize=.55\hsize\leavevmode\epsffile{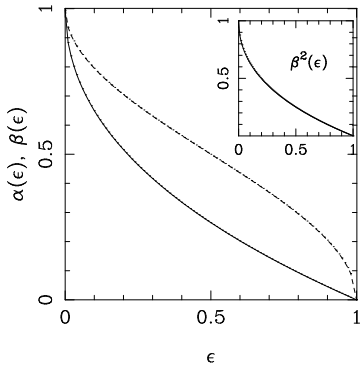}
%\centerline{\epsfig{figure=figure1.ps,width=2.55in}}
\end{center}
\caption{Plot of $\alpha(\epsilon)$ (continuous line) and
$\beta(\epsilon)$ (dashed line) introduced in Eq.~(\ref{newqsl}). The
insert shows the similarity between $\beta^2(\epsilon)$ and
$\alpha(\epsilon)$. }
\labell{f:alpha}\end{figure}
The detailed derivation of Eq.~(\ref{newqsl}) is given in
App.~\ref{s:app}. Here we only give a general idea of the
procedure. The quantity ${\cal T}_\epsilon(E,\Delta E)$ is composed of
two contributions. The first contribution relates the speed of the
dynamical evolution to the average energy $E$ through the function
$\alpha(\epsilon)$ and it extends the Margolus-Levitin theorem
{\cite{margolus}}. It provides the value of ${\cal
T}_\epsilon(E,\Delta E)$ in what we will refer to as the
Margolus-Levitin regime, i.e. when $\Delta
E/E\geqslant\beta(\epsilon)/\alpha(\epsilon)$.  The function
$\alpha(\epsilon)$ is derived by introducing two functions
$\alpha_{\mbox{\tiny{<}}}(\epsilon)$ and
$\alpha_{\mbox{\tiny{>}}}(\epsilon)$ such that
$\alpha_{\mbox{\tiny{<}}}(\epsilon)\leqslant\alpha(\epsilon)\leqslant\alpha_{\mbox{\tiny{>}}}(\epsilon)$. The
first one is obtained by analyzing directly the condition
$P(t)=\epsilon$ and using a class of inequalities that maximizes sines
and cosines with linear functions. The second one is obtained by
studying the time evolution of a class of ``fast'' two level
states. This procedure does not allow one to obtain an explicit
analytical expression for $\alpha(\epsilon)$, however the two bounding
functions $\alpha_<$ and $\alpha_>$ can be shown numerically to
coincide giving an estimate of $\alpha(\epsilon)$. The second
contribution to ${\cal T}_\epsilon(E,\Delta E)$ relates the speed of
the dynamical evolution to the spread $\Delta E$ by means of the
function
\begin{eqnarray}
\beta(\epsilon)=\frac 2\pi\arccos(\sqrt{\epsilon})
\;\labell{betadef}.
\end{eqnarray}
This term provides the value of ${\cal T}_\epsilon(E,\Delta E)$ in
what we will refer to as the Heisenberg regime, i.e. when $\Delta
E/E\leqslant\beta(\epsilon)/\alpha(\epsilon)$.  Equation
(\ref{betadef}) was previously proven in {\cite{bata,uncer}} by
employing the general form of the uncertainty relations. However, for
the sake of completeness, in App.~\ref{s:app} we have rederived the
value of $\beta(\epsilon)$ starting directly from the expression
(\ref{pidit}) of $P(t)$.

Given $E$ and $\Delta E$, the quantum speed limit defines a forbidden
evolution regime where the probability $P(t)$ is not allowed to
enter. In fact, for $0\leqslant t\leqslant{\cal T}_0(E,\Delta E)$,
Eq.~(\ref{newqsl}) implies
\begin{eqnarray}
P(t)\geqslant\max\left\{\alpha^{-1}\left(\frac{2Et}{\pi\hbar}
\right),\beta^{-1}\left(\frac{2\Delta Et}{\pi\hbar}
\right)\right\}
\;\labell{forbid},
\end{eqnarray}
where $\alpha^{-1}$ and $\beta^{-1}$ are the inverse functions of
$\alpha(\epsilon)$ and $\beta(\epsilon)$ respectively. This regime is
shown on Fig.~\ref{f:forbid}, where an example of $P(t)$ trajectory is
plotted. By introducing the Margolus-Levitin type contribution
(i.e. the term dependent on $E$), Eq.~(\ref{forbid}) generalizes the
previous bounds for $P(t)$ {\cite{uncer,bata}}. Notice that, since the
contribution dependent on $\Delta E$ to Eq.~(\ref{forbid}) exhibits a
null derivative in $t=0$, it always provides a non-trivial bound to
$P(t)$. On the other hand, since the contribution dependent on $E$
exhibits a negative slope in $t=0$, it does not provide an achievable
bound when $\Delta E\leqslant E$.  For the same reason, the bound
(\ref{beta2}) on the first derivative of $P(t)$ is not modified by the
presence of the Margolus-Levitin type contribution of
Eq.~(\ref{forbid}): $|dP(t)/dt|$ is limited only by the energy spread
$\Delta E$.

% Da forbidden.f
\begin{figure}[hbt]
\begin{center}
\epsfxsize=.55\hsize\leavevmode
\epsffile{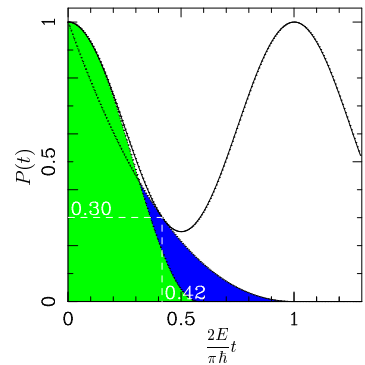}
\end{center}
\caption{Plot of the constraint given by Eq.~(\ref{forbid}), which
shows the forbidden region where $P(t)$ is not allowed to enter
(shaded areas). The time-energy uncertainty contribution to
Eq.~(\ref{forbid}) defines the light grey region through
$\beta^{-1}$. The Margolus-Levitin type contribution defines the dark
grey region through $\alpha^{-1}$.  The condition plotted here is for
$\Delta E/E=1.73$. The continuous line represents the trajectory
$P(t)$ of a ``fast'' state $|\Omega_\xi\rangle$ of Eq.~(\ref{omega})
with $\xi=0.5$. This state touches the boundary of the forbidden
region for $\epsilon=0.30$ and $t={\cal T}_{\epsilon=0.30}(E,\Delta
E)=0.42\;(\frac{\pi\hbar}{2E})$.  }
\labell{f:forbid}\end{figure}

\subsection{Quantum speed limit time for mixed
states}\labell{s:qslmix} Up to now we have focused on pure states of
the system. What happens when the system is in a mixture $\varrho$?
We will show that the notion of quantum speed limit bound
(\ref{newqsl}) can be extended to the density matrices in the sense
that ${\cal T}_\epsilon(E,\Delta E)$ gives the lower bound to the time
it takes for a state $\varrho$ with energy $E$ and spread $\Delta E$
to evolve to a configuration $\varrho(t)$ such that
\begin{eqnarray} {F}(\varrho,\varrho(t))=\epsilon
\;\labell{mixqsl},
\end{eqnarray}
where
${F}(\varrho,\varrho')\equiv\left\{\mbox{Tr}\left[\sqrt{\sqrt{\varrho}{\varrho'}\sqrt{\varrho}}\right]\right\}^2$
is the fidelity introduced in {\cite{fidelity}}.

To prove the above statement, first of all notice that in the case of
pure states, the fidelity reduces to the probability $P(t)$ of
Eq.~(\ref{pidit}) and the definition (\ref{mixqsl}) reduces to the
quantum speed limit bound given in the previous section. More
generally, consider a generic decomposition of $\varrho$,
\begin{eqnarray}
\varrho&=&\sum_np_n|\phi_n\rangle\langle\phi_n|\;\labell{decomp1}
\;\labell{decomp},
\end{eqnarray}
where ${p}_n>0$, $\sum_n{p}_n=1$ and $\{|\phi_n\rangle\}$ is a set
of non-necessarily orthogonal pure states. The fidelity $F$ has
been shown {\cite{fidelity,chuang}} to satisfy the property
\begin{eqnarray}
F(\varrho,\varrho(t))=\max_{{|\chi\rangle,|\chi'\rangle}}
\left\{|\langle\chi|\chi'\rangle|^2\right\}
\;\labell{purif},
\end{eqnarray}
where $|\chi\rangle$ and $|\chi'\rangle$ are purifications of
$\varrho$ and $\varrho(t)$ respectively, such as the states
\begin{eqnarray}
|\chi\rangle&=&\sum_n\sqrt{{p}_n}|\phi_n\rangle|\xi_n\rangle
\;\labell{p1}\\
|\chi'\rangle&=&\sum_n\sqrt{{p}_n}|\phi_n(t)\rangle|\xi'_n\rangle
\;\labell{p},
\end{eqnarray}
with $\{|\xi_n\rangle\}$, $\{|\xi'_n\rangle\}$ being two orthonormal
sets of an ancillary system.  Choosing $|\xi'_n\rangle=|\xi_n\rangle$
for all $n$ and assuming that they are all eigenstates of the ground
level of the ancillary system, $|\chi'\rangle$ becomes the time
evolved of $|\chi\rangle$ (i.e.  $|\chi'\rangle=|\chi(t)\rangle$) and
Eq.~(\ref{purif}) implies that the fidelity is bounded by
\begin{eqnarray}
F(\varrho,\varrho(t))\geq|\langle\chi|\chi(t)\rangle|^2
\;\labell{ca}.
\end{eqnarray}
Since the pure state $|\chi\rangle$ has the same energy $E$ and energy
spread $\Delta E$ of $\varrho$, it can rotate by a quantity $\epsilon$
in a time not smaller than ${\cal T}_\epsilon(E,\Delta E)$, as shown
in the previous section. This, along with inequality (\ref{ca}) proves
that the minimum time $t$ for which $F(\varrho,\varrho(t))=\epsilon$
is bounded by the quantity ${\cal T}_\epsilon(E,\Delta E)$, as stated
in (\ref{mixqsl}).  Notice, finally, that for $\epsilon=0$ we reobtain
all the results that were given in {\cite{role}}, since in this case
the condition (\ref{mixqsl}) is equivalent to the condition
Tr$[\varrho(t)\varrho]=0$ that was employed there {\cite{counter}}.

\subsubsection*{Mixed states that reach the bound}
Before concluding the section, let us analyze under which conditions a
mixed state can reach the quantum speed limit. Assume that the state
$\varrho$ of energy $E$ and spread $\Delta E$ achieves the bound for a
value $\epsilon$, i.e.
\begin{eqnarray}
F\left(\varrho,\varrho\left({\cal T}_\epsilon\right)\right)
=\epsilon
\;\labell{achieve},
\end{eqnarray}
where the dependence on $E$ and $\Delta E$ has been dropped for ease
of notation. Define the quantity
$\epsilon_n\equiv|\langle\phi_n|\phi_n({\cal T}_\epsilon)\rangle|^2$,
which measures the rotation of the $n$-th component of the mixture
(\ref{decomp1}) at time ${\cal T}_\epsilon$. Applying quantum speed
limit considerations to the state $|\phi_n\rangle$, one finds
\begin{eqnarray} 
{\cal T}_\epsilon(E,\Delta E)\geqslant {\cal
T}_{\epsilon_n}({\cal E}_n,\Delta {\cal E}_n)
\;\labell{cb},
\end{eqnarray}
where ${\cal E}_n$ and $\Delta {\cal E}_n$ are the energy and the
spread of the state $|\phi_n\rangle$. Now, among all possible
purifications of the form (\ref{p1}) and (\ref{p}) choose one such
that $\langle\xi_n|\xi'_m\rangle=\delta_{nm}e^{-i\varphi_{nm}}$, where
$\varphi_{nm}=\arg[\langle\phi_n|\phi_m({\cal T}_\epsilon)\rangle]$.
From Eqs.~(\ref{purif}) and (\ref{achieve}), it follows that
\begin{eqnarray}
&{\displaystyle
\epsilon\geqslant\Big|\sum_np_n|\langle\phi_n|\phi_n({\cal
T}_\epsilon)\rangle|\Big|^2=
\Big(\sum_n{p}_n\sqrt{\epsilon_n}\Big)^2
\equiv\bar\epsilon\;,\quad}&
\labell{barep}\\
%\Longrightarrow
&{\cal T}_{\bar\epsilon}(E,\Delta E)\geqslant {\cal
T}_{\epsilon}(E,\Delta E)\;,&
\;\labell{bb}
\end{eqnarray}
where in (\ref{bb}) we employed the fact that $\alpha(\epsilon)$ and
$\beta(\epsilon)$ are strictly decreasing functions.  Combining
Eqs.~(\ref{cb}) and (\ref{bb}), we find that, for all $n$,
\begin{eqnarray}
{\cal T}_{\bar\epsilon}(E,\Delta E)\geqslant{\cal
T}_{\epsilon_n}({\cal E}_n,\Delta {\cal E}_n)
\;\labell{cc}.
\end{eqnarray}
Consider first the Margolus-Levitin regime, i.e. $\Delta E/E
\geqslant\beta(\bar \epsilon)/\alpha(\bar\epsilon)$.  From
Eq.~(\ref{cc}) it follows that
\begin{eqnarray}
\alpha(\bar\epsilon)\frac{\pi\hbar}{2E}\geqslant
\alpha(\epsilon_n)\frac{\pi\hbar}{2{\cal E}_n}
\;\labell{aw}.
\end{eqnarray}
Since the energy of the state $\varrho$ is
$E=\sum_n{p}_n{\cal E}_n$, Eq.~(\ref{aw}) implies \begin{eqnarray}
\alpha(\bar\epsilon)\geqslant\sum_n{p}_n\alpha(\epsilon_n)
\;\labell{azzz}.
\end{eqnarray}
Analogously, in the Heisenberg regime, i.e. when $\Delta E/E \leqslant
\beta(\bar \epsilon)/\alpha(\bar\epsilon)$, since $\Delta
E^2=\sum_n{p}_n[\Delta {\cal E}^2_n+(E-{\cal E}_n)^2]$, one
obtains
\begin{eqnarray}
\beta^2(\bar\epsilon)\geqslant\sum_n{p}_n\beta^2(\epsilon_n)
\;\labell{axxx}.
\end{eqnarray}
The inequalities (\ref{azzz}) and (\ref{axxx}) must be satisfied if
$\varrho$ reaches the quantum speed limit. Since both
$\alpha(\epsilon^2)$ and $\beta^2(\epsilon^2)$ are strictly convex
functions (see Eqs.~(\ref{az}) and (\ref{az1}) of App.~\ref{s:math}),
such conditions can be fulfilled only when the equalities hold: this
happens if $\epsilon_n=\epsilon$ for all $n$ and if the equality holds
also in (\ref{cb}). This shows that the fastest states $\varrho$ are
mixture composed by pure states $|\phi_n\rangle$ that all achieve the
quantum speed limit bound for the same $\epsilon$ at the same time.

% Entanglement
\section{Entangled dynamics}\labell{s:ent}
In a preceding paper {\cite{role}}, we analyzed the role of
entanglement in achieving the quantum speed limit bound (\ref{qsl})
for composite systems. We found that quantum correlations among the
subsystem allow the state of the system to evolve to an orthogonal
configuration faster if the energy resources are not devoted to a
single subsystem. Here we analyze the generalized bound (\ref{newqsl})
and show that the same result holds even when we do not require the
initial and final states to be orthogonal. Quantum correlations among
subsystems allow the state of the system to rotate in Hilbert space
faster if the energy resources are not devoted to a single subsystem.

In the following we consider the case of a composite system with $M$
independent components. Its Hamiltonian is given by $H=\sum_kH_k$,
where $H_k$ is the free Hamiltonian of the $k$-th subsystem. Since the
Hamiltonian $H$ is assumed to have zero ground state, we will redefine
all the $H_k$ to have zero ground states without loss of generality.
We will start analyzing the case of pure states, postponing the case
of non-pure states.

\subsection{Pure states}\labell{s:puro}
Consider a composite system of $M$ non-interacting parts in the
initial pure separable state
\begin{eqnarray} |\Psi\rangle=|\psi_1\rangle_1\cdots|\psi_M\rangle_M
\;\labell{statoin},
\end{eqnarray} which has energy and energy spread 
\begin{eqnarray} E&=&\sum_kE_k\;,\labell{ener}\\
\Delta E&=&\Big(\sum_k\Delta E_k^2\Big)^{1/2}
\;\labell{dener},
\end{eqnarray}
where $E_k$ and $\Delta E_k$ are the energy and the spread of the
state $|\psi_k\rangle_k$ of the $k$-th subsystem {\cite{nota}}.  The
state $|\Psi\rangle$ reaches the quantum speed limit if, for some
value of $\epsilon$, the following identity applies,
\begin{eqnarray} P\left({\cal T}_\epsilon(E,\Delta
E)\right)=\epsilon\;
\labell{condiz},
\end{eqnarray}
where $P(t)$ is the probability (\ref{pidit}) of the state
$|\Psi\rangle$ and ${\cal T}_\epsilon(E,\Delta E)$ is the quantum
speed limit time of Eq.~(\ref{newqsl}). For a separable state, the
quantity $P(t)$ is given by
\begin{eqnarray}
P(t)=P_1(t)\cdots P_M(t)
\;\labell{pixi},
\end{eqnarray}
where $P_k(t)=|_k\langle\psi_k|\psi_k(t)\rangle_k|^2$ is the overlap of
the state of the $k$-th subsystem at time $t$ with its initial
value. Defining $\epsilon_k=P_k({\cal T}_\epsilon(E,\Delta E))$ and
using Eq.~(\ref{pixi}), the condition~(\ref{condiz}) can be rewritten
as
\begin{eqnarray}
\epsilon=\epsilon_1\cdots\epsilon_M\;\labell{equaqua}.
\end{eqnarray}
Moreover, applying the quantum speed limit relation (\ref{newqsl}) to
the $k$-th subsystem, one finds that
\begin{eqnarray} {\cal T}_\epsilon(E,\Delta E)\geqslant {\cal
T}_{\epsilon_k}(E_k,\Delta E_k)
\;\labell{equazione}.
\end{eqnarray}
Consider first the Margolus-Levitin regime, i.e. $\Delta
E/E\geqslant\beta(\epsilon)/\alpha(\epsilon)$. In this case,
Eq.~(\ref{equazione}) and the definitions of ${\cal
T}_\epsilon(E,\Delta E)$ and ${\cal T}_{\epsilon_k}(E_k,\Delta E_k)$
imply
\begin{eqnarray}
\alpha(\epsilon)\frac{\pi\hbar}{2E}\geqslant
\alpha(\epsilon_k)\frac{\pi\hbar}{2E_k}
\;\labell{bellezza},
\end{eqnarray}
and hence, using the expression (\ref{ener}) of the total energy of the
system,
\begin{eqnarray}
\alpha(\epsilon)\geqslant\sum_{k=1}^M\alpha(\epsilon_k)
\;\labell{bellezzissima}.
\end{eqnarray}
Analogously, in the Heisenberg regime Eq.~(\ref{equazione}) implies
\begin{eqnarray}
\beta(\epsilon)\frac{\pi\hbar}{2\Delta E}\geqslant
\beta(\epsilon_k)\frac{\pi\hbar}{2\Delta E_k}
\;\labell{bet},
\end{eqnarray}
and hence
\begin{eqnarray}
\beta^2(\epsilon)\geqslant\sum_{k=1}^M\beta^2(\epsilon_k)
\;\labell{eqsubeta}.
\end{eqnarray}
A necessary condition for the separable state $|\Psi\rangle$ to reach
the quantum speed limit is that there exists a set of $\epsilon_k$
that satisfy at least one of the two inequalities
(\ref{bellezzissima}) or (\ref{eqsubeta}) under the constraint
(\ref{equaqua}). According to the strict subadditivity of
$\alpha(\epsilon)$ and $\beta^2(\epsilon)$ (see Eqs.~(\ref{ax}) and
(\ref{ax1}) of appendix \ref{s:math}), the
relations~(\ref{bellezzissima}) and (\ref{eqsubeta}) can be satisfied
only when the equality holds: this happens if there exists a $k$ (say
$k'$) such that $\epsilon_{k'}=\epsilon$ and $\epsilon_{k}=1$ for
$k\neq k'$.  Such a solution corresponds to the case in which all the
energy resources are devoted to the $k'$-th subsystem. In fact, for
$k=k'$, the relations~(\ref{bellezza}) and (\ref{bet}) imply
\begin{eqnarray}
E_{k'}&\geqslant& \frac{\alpha(\epsilon_{k'})}{\alpha(\epsilon)}
E = E \;
\labell{edik}\\
\Delta E_{k'}&\geqslant& \frac{\beta(\epsilon_{k'})}{\beta(\epsilon)}
\Delta E = \Delta E \;,
\labell{edik1}
\end{eqnarray}
where Eq.~(\ref{edik}) holds in the Margolus-Levitin regime, while
Eq.~(\ref{edik1}) holds in the Heisenberg regime. Equation
(\ref{edik}) and the form (\ref{ener}) of $E$ require that $E_{k'}=E$
and $E_{k}=0$ for $k\neq k'$ {\cite{nota}}.  Since $H_k$ have all zero
ground state energy, this also implies that $\Delta E_{k'}=\Delta E$
and $\Delta E_k=0$ for $k\neq k'$. On the other hand,
Eq.~(\ref{edik1}) and the form (\ref{dener}) of $\Delta E$ require
that $\Delta E_{k'}=\Delta E$ and $\Delta E_k=0$ for $k\neq k'$.

In conclusion, the only states $|\Psi\rangle$ of the form
(\ref{statoin}) that can reach the quantum speed limit (\ref{newqsl})
for some value of $\epsilon$ are those in which all the energy spread
$\Delta E$ is carried by the single subsystem $k'$. The other
subsystems are in eigenstates of their Hamiltonians $H_k$. Moreover,
if the system is in the Margolus-Levitin regime, then $k'$ carries
also all the mean energy $E$, the other subsystems being in their
ground states. From the dynamical point of view, this means that $k'$
is the only subsystem that rotates in the Hilbert space, while all the
others do not evolve.

\subsubsection*{A simple example: separable vs. entangled
state}\labell{s:homog} The gap between entangled states and non
entangled ones is particularly evident is the case in which the energy
resources are homogeneously distributed among all subsystems,
i.e. when $E_k=E/M$ and $\Delta E_k=\Delta E/\sqrt{M}$ for all
$k$. For the sake of simplicity we analyze an example in which all the
subsystems are in the same state $|\psi_k\rangle_k=|\psi\rangle_k$. In
this case the minimum time $t_\epsilon$ for which the global state
$|\Psi_s\rangle=|\psi\rangle_1\cdots|\psi\rangle_M$ rotates by a
quantity $\epsilon$ is given by the minimum time it takes for each
subsystem to rotate by a quantity $\epsilon^{1/M}$. Applying the
quantum speed limit (\ref{newqsl}) to each subsystem, one obtains that
$t_\epsilon\geqslant{\cal T}_{\epsilon^{1/M}}(E/M,\Delta E/\sqrt{M})$.
The ratio $R(\epsilon)$ between $t_\epsilon$ and ${\cal
T}_\epsilon(E,\Delta E)$, i.e. \begin{eqnarray}
R(\epsilon)\geqslant\min\Big(
{M}\frac{\alpha(\epsilon^{1/M})}{\alpha(\epsilon)},
\sqrt{M}\frac{\beta(\epsilon^{1/M})}{\beta(\epsilon)}\Big)
\;\labell{rat},
\end{eqnarray}
shows how much slower the separable state $|\Psi_s\rangle$ is in
comparison with the maximum speed allowed for a system of the same
energetic resources of $|\Psi_s\rangle$ (see Fig.~\ref{f:ratio}). From
the subadditivity properties (\ref{ax}) and (\ref{ax1}), it follows
that $R(\epsilon)\geqslant 1$. In particular, for $\epsilon=0$,
$R(\epsilon)$ is always greater than $\sqrt{M}$, as discussed in
{\cite{role}}.

% da ratio.f
\begin{figure}[hbt]
\begin{center}
\epsfxsize=.75\hsize\leavevmode
\epsffile{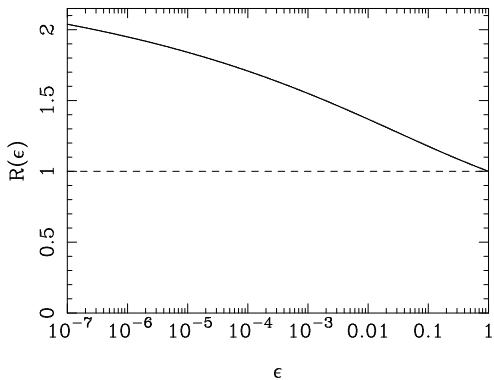}
\end{center}
\caption{Plot of the lower bound of $R(\epsilon)$ from
Eq.~(\ref{rat}). It shows that homogeneous separable states cannot
achieve the quantum speed limit bound (given by the dashed line). Here
$M=5$. }
\labell{f:ratio}\end{figure}

Consider now the following family of entangled states \begin{eqnarray}
|\Psi_\xi\rangle=\sqrt{1-\xi^2}\;|0\rangle_1\cdots|0\rangle_M+{\xi}\;
|E_0\rangle_1\cdots|E_0\rangle_M
\;\labell{omegaent},
\end{eqnarray} 
where $\xi\in[0,1]$ and $|0\rangle$ and $|E_0\rangle$ are eigenstates
of the Hamiltonian of energy $0$ and $E_0>0$ respectively. The state
$|\Psi_\xi\rangle$ represents an homogeneous configuration where each
subsystem has energy $\xi^2E_0$ and spread
$\xi^2\sqrt{1-\xi^2}E_0$. However, unlike the separable state
$|\Psi_s\rangle$ discussed before, for a suitable choice of the
parameter $\xi$, $|\Psi_\xi\rangle$ achieves the quantum speed limit
bound, as can be shown by comparison with the state
$|\Omega_\xi\rangle$ of Eq.~(\ref{omega}) in App.~\ref{s:alpha}.

Proving that homogeneous separable states cannot exhibit speedup while
at least one homogeneous entangled case that exhibits speedup exists,
we have shown that entanglement is a fundamental resource in this
context.

\subsection{Entangled dynamics for mixed states}\labell{s:mix}
In this section we generalize the results of the previous section to
mixed states.

The most general separable state of $M$ subsystems has the form 
\begin{eqnarray}
\varrho=\sum_np_n|\Psi^{(n)}\rangle\langle\Psi^{(n)}|
\;\labell{sm},
\end{eqnarray}
where $p_n>0$, $\sum_np_n=1$, and
\begin{eqnarray}
|\Psi^{(n)}\rangle=|\psi^{(n)}_1\rangle_1\cdots
|\psi^{(n)}_M\rangle_M
\;\labell{sm1},
\end{eqnarray}
with $|\psi^{(n)}_k\rangle_k$ a state of the $k$-th subsystem.  As
discussed in Sect.~\ref{s:qslmix}, ${\cal T}_\epsilon(E,\Delta E)$ is
the minimum time it takes for the state $\varrho$ to reach a
configuration $\varrho(t)$ with fidelity $\epsilon$. Moreover, we
already know that the state $\varrho$ will reach the quantum speed
limit bound for mixed states of Eq.~(\ref{mixqsl}) only if all the
states $|\Psi^{(n)}\rangle$ rotate by an amount $\epsilon$ in the time
${\cal T}_\epsilon({\cal E}_n,\Delta {\cal E}_n)$, given ${\cal E}_n$
and $\Delta {\cal E}_n$ the energy and energy spread of
$|\Psi^{(n)}\rangle$. Since $|\Psi^{(n)}\rangle$ is a separable pure
state, from the previous section it follows that this is possible only
if there exist a subsystem (say the $k_n$-th) that possesses all the
energy resources. This means that the only separable states $\varrho$
that reach the bound (\ref{mixqsl}) are those for which for any
statistical realization $n$ of the mixture (\ref{sm}), a single
subsystem evolves to an orthogonal configuration at its own maximum
speed limit time (which coincides with ${\cal T}(E,\Delta E)$ of the
whole system). All the other subsystems do not evolve. Since the above
derivation applies for any expansion $p_n$, one can say that in each
experimental run only one of the subsystems evolves. This is
essentially the same result that was obtained in {\cite{role}},
although here we considered the more general case of $\epsilon\neq 0$.

% Conclusione
\section{Conclusion}\labell{s:concl}
In this paper we have generalized the definition of quantum speed
limit time {\cite{margolus,role}} to take into account the case in
which the system does not evolve to an orthogonal state. We have used
the fidelity $F(\varrho,\varrho(t))$ of {\cite{fidelity}} as a measure
of the ``distance'' between the initial and the final states. In this
context we have analyzed the role of quantum correlations among
subsystems in a composite system. As a result we have shown that
entanglement is a fundamental resource to achieve a speedup in the
dynamical evolution in a composite system. In fact, the only separable
states that can achieve the quantum speed limit bound are those where
only one subsystem at a time is evolving, while the others are
stationary.

\appendix\section{}\labell{s:app}
In Sects. \ref{s:alpha} and \ref{s:beta} we derive the form of the
functions $\alpha(\epsilon)$ and $\beta(\epsilon)$ respectively. In
Sect.~\ref{s:math} we study these two functions giving some
mathematical properties that are used in the paper.

\subsection{Derivation of $\alpha(\epsilon)$.}\labell{s:alpha}
In order to determine $\alpha(\epsilon)$ we will: {\bf{\em i)}} give a
lower bound for it; {\bf{\em ii)}} give an upper bound for it;
{\bf{\em iii)}} show numerically that these two bounds coincide, thus
providing an estimation of $\alpha(\epsilon)$.

{\bf{\em i)}} A lower bound for $\alpha(\epsilon)$ can be obtained by
observing that if $P(t)=\epsilon$, then
$\langle\Psi|\Psi(t)\rangle=\sqrt{\epsilon}\;e^{i\theta}$, i.e.
\begin{eqnarray}
\sum_n|c_n|^2\cos\frac{E_nt}\hbar&=&\sqrt{\epsilon}\cos\theta
\nonumber\\
\sum_n|c_n|^2\sin\frac{E_nt}\hbar&=&-\sqrt{\epsilon}\sin\theta
\;\labell{part1},
\end{eqnarray}
with $\theta\in[0,2\pi]$.  Consider now the following class of
inequalities for $q\geqslant 0$:
\begin{eqnarray}
\cos x+q\sin x\geqslant 1-ax,
\;\labell{disug}
\end{eqnarray}
which is valid for $x\geqslant 0$ and where $a$ is a function of $q$
defined implicitly by the set of equations
\begin{eqnarray}
\left\{\begin{array}{ccl}
 a&=&\displaystyle\frac{y+\sqrt{y^2(1+q^2)+q^2}}{1+y^2}\cr\cr \sin
 y&=&\displaystyle\frac{a(1-qy)+q}{1+q^2}\;,\end{array}\right.
\;\labell{adiqdef}
\end{eqnarray}
for $y\in[\pi-\arctan(1/q),\pi+\arctan(q)]$.  The inequality
(\ref{disug}) is obtained by bounding the term on the left with the
linear function that is tangent to it and is equal to 1 for $x=0$, as
shown in Fig.~\ref{f:tang}.
% Da maggiorazione.f
\begin{figure}[hbt]
\begin{center}
\epsfxsize=.55\hsize\leavevmode
\epsffile{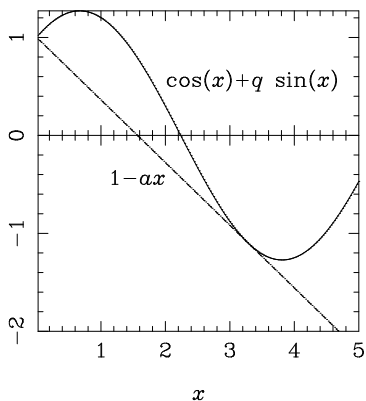}
\end{center}
\caption{Plot of the inequality (\ref{disug}) for $q=\frac\pi4$. In
this case $a\simeq.64$.}
\labell{f:tang}\end{figure}
Since we assumed zero ground state energy, all the energy levels are
positive and we can replace $x$ with $E_nt/\hbar$ in
Eq.~(\ref{disug}). Summing on $n$ and employing Eq.~(\ref{part1}), we
obtain the inequality
\begin{eqnarray}
\sqrt{\epsilon}\left(\cos\theta-q\sin\theta\right)\geqslant
1-a\frac{Et}\hbar
\;\labell{bellissima}.
\end{eqnarray}
From the definition of $\alpha(\epsilon)$ introduced in
Eq.~(\ref{newqsl}), this implies
\begin{eqnarray}
\alpha(\epsilon)\geqslant
[1-\sqrt{\epsilon}\left(\cos\theta-q\sin\theta\right)]
\frac 2{\pi a}
\;\labell{part2}.
\end{eqnarray}
{\comment{aggiungere figura con l'envelope della condizione
(\ref{part2}).}}
Since, for a given value of $\theta$, Eq.~(\ref{part2}) must be valid
for all $q\geqslant 0$, then the following lower bound for
$\alpha(\epsilon)$ can be obtained \begin{eqnarray}
&&\alpha(\epsilon)\geqslant\alpha_{\mbox{\tiny{<}}}(\epsilon)
\nonumber\\
&&\equiv
\min_\theta\left\{\max_q\left\{
[1-\sqrt{\epsilon}\left(\cos\theta-q\sin\theta\right)]
\frac 2{\pi a}\right\}\right\}
\;\labell{part3}.
\end{eqnarray}

{\bf{\em ii)}} To provide an upper bound for $\alpha(\epsilon)$
consider the following family of two-level states,
\begin{eqnarray} |\Omega_\xi\rangle=\sqrt{1-\xi^2}\;|0\rangle+{\xi}\;
|E_0\rangle
\;\labell{omega},
\end{eqnarray} 
where $\xi\in[0,1]$, and $|0\rangle$ and $|E_0\rangle$ are
eigenstates of the Hamiltonian of energy $0$ and $E_0$ respectively.
The state $|\Omega_\xi\rangle$ has average energy $E=\xi^2E_0$
and energy spread $\Delta E=\xi\sqrt{1-\xi^2}E_0$.
Solving the dynamical evolution of the state $|\Omega_\xi\rangle$,
one can show that the first time $t$ for which $P(t)=\epsilon$ is
given by \begin{eqnarray}
\frac{Et}{\hbar}=\xi^2\arccos
\left[\frac{\epsilon-1+2\xi^2(1-\xi^2)}
{2\xi^2(1-\xi^2)}\right]
\;\labell{timet}.
\end{eqnarray}
Minimizing over $\xi$ the right term of Eq.~(\ref{timet}) gives the
following upper bound for $\alpha(\epsilon)$, i.e.
\begin{eqnarray}
\alpha(\epsilon)\leqslant\alpha_{\mbox{\tiny{>}}}(\epsilon)\equiv
\frac 2\pi\; z\;\arccos\left[
\frac{\epsilon-1+2z(1-z)}{2z(1-z)}\right]
\;\labell{pippo},
\end{eqnarray}
where the $z$ is a function of $\epsilon$ defined implicitly by 
\begin{eqnarray}
&&\arccos\left[
\frac{\epsilon-1+2z(1-z)}{2z(1-z)}\right]\nonumber\\&&=
\frac{1-2z}{1-z}\sqrt{\frac{1-\epsilon}{\epsilon-1+4z(1-z)}}
\;\labell{pippo2}.
\end{eqnarray}

{\bf{\em iii)}} The obvious difficulty in deriving the explicit analitic
form of the bounds $\alpha_{\mbox{\tiny{<}}}(\epsilon)$ and
$\alpha_{\mbox{\tiny{>}}}(\epsilon)$ defined in (\ref{part3}) and
(\ref{pippo}), can be overcome by performing a numerical study of
these two conditions. We will show that
$\alpha_<(\epsilon)=\alpha_>(\epsilon)$, thus giving an estimate of
$\alpha(\epsilon)$.

%da dimostraz.f
\begin{figure}[hbt]
\begin{center}
\epsfxsize=.55\hsize\leavevmode\epsffile{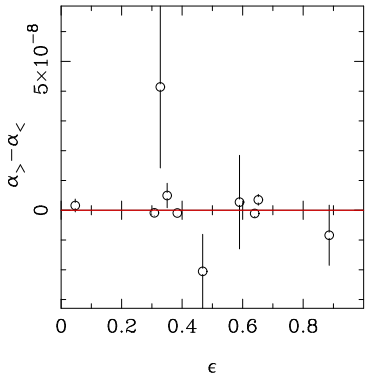}
\end{center}
\vspace{-.5cm}
\caption{Comparison of $\alpha_<(\epsilon)$ and
$\alpha_>(\epsilon)$ for some random values of $\epsilon$. The plot
shows $\alpha_>(\epsilon)-\alpha_<(\epsilon)$ with the error bars
denoting the $\delta\theta\to 0$ extrapolation error of
$\alpha_<(\epsilon)$ (see text). Since all the values are compatible
with zero, we can conclude that
$\alpha_<(\epsilon)=\alpha_>(\epsilon)=\alpha(\epsilon)$.}
\labell{f:dimost}\end{figure}

In order to numerically estimate $\alpha_>(\epsilon)$ one has to solve
Eq.~(\ref{pippo2}). Using a bisection algorithm, it is simple to get a
machine-precision accurate solution very rapidly for all values of
$\epsilon$. On the other hand, the estimate of $\alpha_<(\epsilon)$
requires greater care, since two different parameters ---$q$ and
$\theta$ of Eq.~(\ref{part3})--- are present in its definition. For
each value of $\epsilon$ it is necessary to calculate the term on the
right of Eq.~(\ref{part2}) on a bidimensional grid of values of $q$
and $\theta$ and find for each $\theta$ the maximum on $q$. The value
of $\alpha_<(\epsilon)$ is calculated by choosing the minimum among
these maxima. Of course this procedure is biased, since it depends on
the grid spacings $\delta q$ and $\delta\theta$. In order to remove
the bias in the calculation result, one can repeat the whole procedure
for different values of the grid spacing and then extrapolate the
result for the spacings $\delta q$ and $\delta\theta$ tending to
zero. We have used a least square linear interpolation, where $\chi^2$
minimization allows also to recover an ``error bar'' that measures how
well the linear interpolation works for each value of $\epsilon$.  The
error bar has no statistical meaning: it simply gives an idea of how
well the linear extrapolation for $\delta\theta\to 0$ works for the
value of $\epsilon$ under consideration. It can be used also to give a
``confidence interval'' for the obtained result.  The $\delta q\to 0$
extrapolation error has been found negligible in all cases, meaning
that a linear extrapolation is well suited. To reduce aliasing
problems, instead of using an equispaced grid, it is preferable to
adopt a random grid of values of $q$ and $\theta$ uniformly
distributed so that the average distance between distinct values is
$\delta q$ and $\delta\theta$ respectively. {\comment{In particular,
the algorithm steps for the evaluation of $\alpha_<(\epsilon)$ are:
1)~Define a random uniformly-spaced grid of values of $\theta$ with
average spacing $\delta\theta$; 2)~For each $\theta$ define a number
$N_q$ of random uniformly-spaced grids of values of $q$ with average
spacing $\delta q$, $\delta q/2$, ... , $\delta q/2^{N_q}$; 3)~For
each $q$ of the $N_q$ grids solve Eq.~(\ref{adiqdef}) with a bisection
algorithm; 4)~For each of the $N_q$ grids calculate
$\max_q\left\{[1-\sqrt{\epsilon}\left(\cos\theta-q\sin\theta\right)]\frac
2{\pi a}\right\}$; 5)~Extrapolate such values for $\delta q\to 0$
checking that the extrapotion ``error bar'' (estimated using the
minimization of $\chi^2$) is negligible; 6)~Find the minimum over
$\theta$ of the values obtained in 5) and repeat the whole procedure
for $N_\theta$ different grids of values of $\theta$ with decreasing
average spacing; 7)~Extrapolate the results to $\delta\theta\to 0$,
estimating (through $\chi^2$ minimization) the ``error bar'' of the
result.}}

The extrapolated value of $\alpha_<(\epsilon)$ with its error bars is
compared with the calculated value of $\alpha_>(\epsilon)$ in
Fig.~\ref{f:dimost}.  Machine-precision accuracy is rapidly attainable
in the calculation of $\alpha_>(\epsilon)$ and we have considered it
as unaffected by error. Since the values of $\alpha_>(\epsilon)$ and
$\alpha_<(\epsilon)$ are compatible for arbitrary values of $\epsilon$
we can conclude that the two functions coincide and are thus both
equal to $\alpha(\epsilon)$. This allows to give the numerical
estimations of this function that have been used throughout the
paper. Notice, however, that $\alpha(\epsilon)$ is roughly
approximated (up to a few percent error) by the function
$\beta^2(\epsilon)$ as can be seen from Fig.~\ref{f:alpha}.

\subsection{Derivation of $\beta(\epsilon)$.}\labell{s:beta}
The function $\beta(\epsilon)$ can be derived starting from
Eq.~(\ref{pidit}) by the following chain of relations:
\begin{widetext}
\begin{eqnarray}
\left|\frac {d }{dt}P(t)\right|&=&\frac 2\hbar
\left|\sum_{n,m}|c_n|^2|c_m|^2(E_n-E)\sin\left(\frac{E_n-E_m}\hbar
t\right)\right|%\nonumber\\&\leqslant&
\leqslant\frac 2\hbar
\left|\sum_{n,m}|c_n|^2|c_m|^2(E_n-E)e^{-i(E_n-E_m)
t/\hbar}\right|\nonumber\\&=&
\frac 2\hbar
\left|\sum_{n}|c_n|^2(E_n-E)\left(\sum_m|c_m|^2e^{-i(E_n-E_m)
t/\hbar}-P(t)\right)\right|
\;\labell{beta},
\end{eqnarray}
\end{widetext}
where the last identity has been obtained adding a zero term to the
sum on $n$. Applying the Cauchy-Schwarz inequality to
Eq.~(\ref{beta}), we find \begin{eqnarray}
\left|\frac {d }{dt}P(t)\right|\leqslant \frac{2\Delta
E}\hbar\sqrt{P(t)[1-P(t)]} 
\;\labell{beta2},
\end{eqnarray}
which for $0\leqslant t\leqslant \pi\hbar/(2\Delta E)$ implies
{\cite{bata,uncer}}
\begin{eqnarray}
P(t)\geqslant\cos^2\left(\frac{\Delta E}\hbar t\right)
\;\labell{beta3}.
\end{eqnarray}
This means that the smallest time $t$ for which $P(t)=\epsilon$ is
bounded by the quantity $\beta(\epsilon)\pi\hbar/(2\Delta E)$ with
$\beta(\epsilon)$ defined in (\ref{betadef}).  Notice that the bound
(\ref{beta3}) is achievable since, for example, the state
$|\Omega_{\xi=1/\sqrt{2}}\rangle$ of Eq.~(\ref{omega}) reaches it.

\subsection{Mathematical properties of $\alpha(\epsilon)$ and
$\beta(\epsilon)$}\labell{s:math} 

Both $\alpha(\epsilon)$ and $\beta(\epsilon)$ are strictly decreasing
functions (see Fig.~\ref{f:alpha}). Moreover they satisfy the
following constraints:\begin{enumerate}\item[{\bf{\em a)}}]\label{ita}
The functions $\alpha(\epsilon^2)$ and $\beta^2(\epsilon^2)$ are
strictly convex, i.e. for $\epsilon_n\in[0,1]$,\begin{eqnarray}
\alpha\Big(\Big(\sum_{n=1}^N{p}_n\epsilon_n\Big)^2\Big)
&\leqslant&\sum_{n=1}^N{p}_n\alpha(\epsilon_n^2)
\;\labell{az},\\
\beta^2\Big(\Big(\sum_{n=1}^N{p}_n\epsilon_n\Big)^2\Big)
&\leqslant&\sum_{n=1}^N{p}_n\beta^2(\epsilon_n^2)
\;\labell{az1},
\end{eqnarray}
where ${p}_n>0$ and $\sum_n{p}_n=1$. The identity in
(\ref{az}) and (\ref{ax}) holds only if $\epsilon_n=\epsilon_{n'}$ for
all $n$ and $n'$.\item[{\bf{\em b)}}\label{itb}] The functions
$\alpha(\epsilon)$ and $\beta^2(\epsilon)$ are strictly subadditive,
i.e. for $\epsilon_k\in[0,1]$,\begin{eqnarray}
\alpha\Big(\prod_{k=1}^N\epsilon_k\Big)
&\leqslant&\sum_{k=1}^N\alpha(\epsilon_k)
\;\labell{ax},\\
\beta^2\Big(\prod_{k=1}^N\epsilon_k\Big)
&\leqslant&\sum_{k=1}^N\beta^2(\epsilon_k)
\;\labell{ax1},
\end{eqnarray}
with the identity holding only when there exists a $k$ (say $k'$) such
that $\epsilon_k=1$ for all $k\neq k'$.
\end{enumerate}
To prove these properties, one can discuss the case of $N=2$ and then
extend it by induction to arbitrary $N$.  When referred to
$\beta(\epsilon)$, both propertires can be analitically proved using
its definition (\ref{betadef}). For $\alpha(\epsilon)$ we must instead
resort to numerical verification (e.g. see Fig.~\ref{f:soluz} and
\ref{f:soluz1}).

% Da soluz5.f
\begin{figure}[hbt]
\begin{center}
\epsfxsize=.85\hsize\leavevmode\epsffile{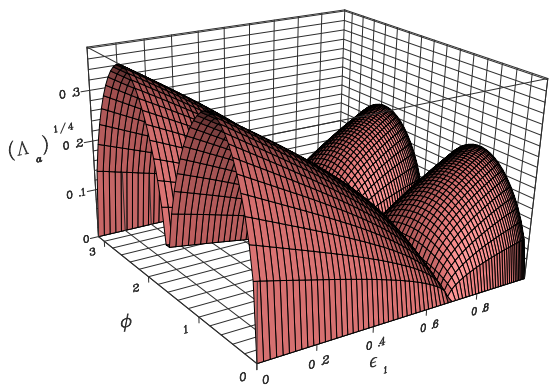}
\end{center}
\vspace{-.5cm}
\caption{Convexity condition (\ref{az}) for $\alpha(\epsilon^2)$ in the
case $N=2$:
$\Lambda_a\equiv\alpha(\epsilon_1^2)\cos^2\phi+\alpha(\epsilon_2^2)\sin^2\phi-\alpha((\epsilon_1\cos^2\phi+\epsilon_2\sin^2\phi)^2)\geqslant
0$. In this plot $\epsilon_2=0.7$. Notice that $\Lambda_a$ is null
only for $\epsilon_1=\epsilon_2$ and for $\phi=0,\pi$.}
\labell{f:soluz}\end{figure}

% Da soluz2.f
\begin{figure}[hbt]
\begin{center}
\epsfxsize=.85\hsize\leavevmode\epsffile{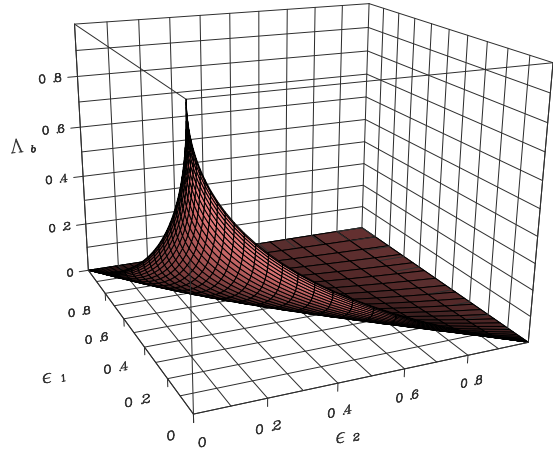}
\end{center}
\vspace{-.5cm}
\caption{Subadditivity condition (\ref{ax}) for $\alpha(\epsilon)$
in the case $N=2$:
$\Lambda_b\equiv\alpha(\epsilon_1)+\alpha(\epsilon_2)-\alpha(\epsilon_1\epsilon_2)\geqslant
0$, for $\epsilon_1,\epsilon_2\geqslant 0$. Notice that $\Lambda_b$ is
null only for $\epsilon_1=1$ or $\epsilon_2=1$. }
\labell{f:soluz1}\end{figure}

%\begin{acknowledgments}
This work was funded by the ARDA, NRO, and by ARO under a MURI
program.
%\end{acknowledgments}


\begin{references}
\bibitem{counter} One is tempted to interpret the time ${\cal
T}_\epsilon(E,\Delta E)$ as the minimum time it takes for the system
to evolve to a configuration $\varrho'$ such that
$\mbox{Tr}[\varrho'\varrho]=\epsilon\mbox{Tr}[\varrho^2]$ (where the
term Tr$[\varrho^2]$ is needed for normalization purposes). However,
it is possible to find counterexamples that show that this definition
does not work when $\epsilon\neq 0$. For example consider the mixture
\begin{eqnarray}
\varrho&=&\cos^2\phi\;|\chi_1\rangle\langle\chi_1|+
\sin^2\phi\;|\chi_2\rangle\langle\chi_2|
\;\nonumber,
\end{eqnarray}
where
$|\chi_{1,2}\rangle=\left(|0\rangle+|E_{1,2}\rangle\right)|\zeta_{1,2}\rangle/\sqrt{2}$,
with $|0\rangle$ and $|E_{1,2}\rangle$ energy eigenstates and
$|\zeta_{1,2}\rangle$ ancillary orthogonal states with null
energy. This state will reach the condition
$\mbox{Tr}[\varrho(t)\varrho]=\epsilon\mbox{Tr}[\varrho^2]$ in a time
$t=0.400\hbar/E_1$ smaller than ${\cal T}_\epsilon(E,\Delta
E)=0.410\hbar/E_1$ if the following parameters are employed:
$\epsilon=0.2$, $\phi=1.070$, $E_2/E_1=1.425$.
\bibitem{nota} Since all the $H_k$ have zero energy ground
states, then $E_k\geqslant 0$ for all $k$.
\end{references}
\end{document}